\documentclass[10pt,twoside]{article}

\usepackage{amsfonts}
\usepackage{amssymb}
\usepackage{amsmath}
\usepackage{pifont}
\usepackage{graphicx}
\usepackage[english]{babel}

\numberwithin{equation}{section}

\begin{document}

\title{$331$ vector-like models with mirror fermions as a possible solution
for the discrepancy in the $b-$quark asymmetries, and for the neutrino mass
and mixing pattern}
\author{Rodolfo A. Diaz\thanks{%
radiazs@unal.edu.co}, R. Martinez\thanks{%
remartinezm@unal.edu.co}, F. Ochoa\thanks{%
faochoap@unal.edu.co} \\
%EndAName
Universidad Nacional de Colombia, \\
Departamento de F\'{\i}sica. Bogot\'{a}, Colombia.}
\date{}
\maketitle

\vspace{-8mm}

\begin{abstract}
A general study of the fermionic structure of the 331 models with $\beta $
arbitrary shows the possibility of obtaining 331 vector-like models with
mirror fermions. On one hand, the existence of mirror fermions gives a
possible way to fit the discrepancy in the bottom quark asymmetries from the
prediction of the standard model. On the other hand, the vector-like nature
of the model permits to address the problem of the fermion mass hierarchy,
and in particular the problem of the neutrino mass and mixing pattern.
Specifically, we consider a model with four families and $\beta =-1/\sqrt{3}$
where the additional family corresponds to a mirror fermion of the third
generation of the Standard Model. We also show how to generate ansatzs about
the mass matrices of the fermions according to the phenomenology. In
particular, it is possible to get a natural fit for the neutrino
hierarchical masses and mixing angles. Moreover, by means of the mixing
between the third quark family and its mirror fermion, a possible solution
for the $A_{FB}^{b}$ discrepancy is obtained.


\end{abstract}

\vspace{-8mm}

\section{Introduction}

The models with gauge symmetry $SU(3)_{c}\otimes SU(3)_{L}\otimes U(1)_{X},$
also called 331 models are well motivated models that could address problems
such as the origin of families, the hierarchy problem in grand unified
theories and the charge quantization problem \cite{fourteen}. Nevertheless,
current versions of the 331 model cannot provide an explanation about the
mass hierarchy and mixing of the fermions, because models with vector-like
multiplets are necessary to explain the family hierarchy. In particular, the
neutrinos do not exhibit a strong family hierarchy pattern as it happens
with the other fermions; the mixing angles for the atmospheric and solar
neutrinos\ are not small; and the quotient $\left( \delta m_{sun}^{2}/\delta
m_{atm}^{2}\right) $ is of the order of $0.02-0.03$, these facts suggest to
modify the see-saw mechanism in order to cancel the hierarchy in the mass
generation for the neutrinos, such modifications are usually implemented by
introducing vector-like fermion multiplets \cite{Valle}. On the other hand,
since the traditional 331 models are purely left-handed, they cannot account
about parity breaking. Moreover, their left-handed nature along with the
weakness of the $Z-Z^{\prime }$ mixing, prevent such models to explain the
deviation of the $b-$quark asymmetry $A_{b}$ from the value predicted by SM 
\cite{Martinez}. An interesting alternative to solve this puzzle is to
consider models with mirror fermions (MF) that couple with right-handed
chirality to the electroweak gauge fields from which the $A_{b}$ and $%
A_{FB}^{b}$ deviations could be fitted \cite{Chanowitz}. The fitting of this
deviation could be achieved by either a modification in the right-handed
couplings of $Z_{\mu }$ with the $b-$quark or by modifying the right-handed
couplings of the top quark which enter in the correction of the $Zb\overline{%
b}$ vertex. As a consequence, the $\left\vert V_{tb}\right\vert $ CKM
element could change as well, this fact could give a hint concerning the
mass generation mechanism for the ordinary fermions.

According to the discussion above, it would be desirable to get vector-like
331 models with mirror fermions, that might in principle be able to generate
right-handed couplings for the bottom and top quarks and at the same time
provide a possible explanation to the fermion mass hierarchy. A recent study
shows the possibility of finding 331 models with such features \cite%
{331mirror}. The present manuscript describes the minimal 331 vector-like
model with MF obtained in Ref. \cite{331mirror} and the way in which such
model could solve the problems addressed above.

\vspace{-4mm}

\section{General fermionic structure}

In the 331 models, the electric charge is defined as a linear combination of
the diagonal generators of the group 
\begin{equation}
Q=T_{3}+\beta T_{8}+XI,
\end{equation}%
and the value of the $\beta $ parameter determines the fermion assignment
and the electric charges of the exotic spectrum \cite{fourteen, 331us}.
Hence, such parameter is used to classify the different 331 models. An
analysis of the fermion representations shows that the left-handed
multiplets lie in either the $\mathbf{3}$ or $\mathbf{3}^{\ast }$
representations of $SU\left( 3\right) _{L}$.

Ref. \cite{331mirror} has studied the possible fermionic structures for 331
models with $\beta $ arbitrary based on cancellation of anomalies and
demanding a fermionic spectrum with a minimal number of exotic particles.
Denoting $N\ $as the number of leptonic generations and $M$ the number of
quark generations, minimization of the exotic fermionic spectrum requires to
associate only one lepton and one quark $SU\left( 3\right) _{L}\ $multiplet
with each generation, and at most one right-handed singlet associated with
each left-handed fermion. From such assumptions we obtain the fermionic
spectrum (containing the SM spectrum) displayed in table \ref{tab:fercont}
for the quarks and leptons. 
\begin{table}[!tbp]
\begin{center}
\begin{equation*}
\begin{tabular}{||c||c||c||}
\hline\hline
$Quarks$ & $Q_{\psi }$ & $X_{\psi }$ \\ \hline\hline
\begin{tabular}{c}
$q_{L}^{(m)}=\left( 
\begin{array}{c}
U^{(m)} \\ 
D^{(m)} \\ 
J^{(m)}%
\end{array}
\right) _{L}:\mathbf{3}$ \\ 
\\ 
$U_{R}^{(m)}:\mathbf{1}$ \\ 
$D_{R}^{(m)}:\mathbf{1}$ \\ 
$J_{R}^{(m)}:\mathbf{1}$%
\end{tabular}
& 
\begin{tabular}{c}
$\left( 
\begin{array}{c}
\frac{2}{3} \\ 
-\frac{1}{3} \\ 
\frac{1}{6}-\frac{\sqrt{3}\beta }{2}%
\end{array}
\right) $ \\ 
\\ 
$\frac{2}{3}$ \\ 
$-\frac{1}{3}$ \\ 
$\frac{1}{6}-\frac{\sqrt{3}\beta }{2}$%
\end{tabular}
& 
\begin{tabular}{c}
\\ 
$X_{q^{(m)}}^{L}=\frac{1}{6}-\frac{\beta }{2\sqrt{3}}$ \\ 
\\ 
\\ 
$X_{U^{(m)}}^{R}=\frac{2}{3}$ \\ 
$X_{D^{(m)}}^{R}=-\frac{1}{3}$ \\ 
$X_{J^{(m)}}^{R}=\frac{1}{6}-\frac{\sqrt{3}\beta }{2}$%
\end{tabular}
\\ \hline\hline
\begin{tabular}{c}
$q_{L}^{(m^{\ast })}=\left( 
\begin{array}{c}
D^{(m^{\ast })} \\ 
-U^{(m^{\ast })} \\ 
J^{(m^{\ast })}%
\end{array}
\right) _{L}:\mathbf{3}^{\ast }$ \\ 
\\ 
$D_{R}^{(m^{\ast })}:\mathbf{1}$ \\ 
$U_{R}^{(m^{\ast })}:\mathbf{1}$ \\ 
$J_{R}^{(m^{\ast })}:\mathbf{1}$%
\end{tabular}
& 
\begin{tabular}{c}
$\left( 
\begin{array}{c}
-\frac{1}{3} \\ 
\frac{2}{3} \\ 
\frac{1}{6}+\frac{\sqrt{3}\beta }{2}%
\end{array}
\right) $ \\ 
\\ 
$-\frac{1}{3}$ \\ 
$\frac{2}{3}$ \\ 
$\frac{1}{6}+\frac{\sqrt{3}\beta }{2}$%
\end{tabular}
& 
\begin{tabular}{c}
\\ 
$X_{q^{(m^{\ast })}}^{L}=-\frac{1}{6}-\frac{\beta }{2\sqrt{3}}$ \\ 
\\ 
\\ 
$X_{D^{(m^{\ast })}}^{R}=-\frac{1}{3}$ \\ 
$X_{U^{(m^{\ast })}}^{R}=\frac{2}{3}$ \\ 
$X_{J^{(m^{\ast })}}^{R}=\frac{1}{6}+\frac{\sqrt{3}\beta }{2}$%
\end{tabular}
\\ \hline\hline
$Leptons$ & $Q_{\psi }$ & $X_{\psi }$ \\ \hline\hline
\begin{tabular}{c}
$\ell _{L}^{(n)}=\left( 
\begin{array}{c}
\nu ^{(n)} \\ 
e^{(n)} \\ 
E^{(n)}%
\end{array}
\right) _{L}:\mathbf{3}$ \\ 
\\ 
$\nu _{R}^{(n)}:\mathbf{1}$ \\ 
$e_{R}^{(n)}:\mathbf{1}$ \\ 
$E_{R}^{(n)}:\mathbf{1}$%
\end{tabular}
& 
\begin{tabular}{c}
$\left( 
\begin{array}{c}
0 \\ 
-1 \\ 
-\frac{1}{2}-\frac{\sqrt{3}\beta }{2}%
\end{array}
\right) $ \\ 
\\ 
$0$ \\ 
$-1$ \\ 
$-\frac{1}{2}-\frac{\sqrt{3}\beta }{2}$%
\end{tabular}
& 
\begin{tabular}{c}
\\ 
$X_{\ell ^{(n)}}^{L}=-\frac{1}{2}-\frac{\beta }{2\sqrt{3}}$ \\ 
\\ 
\\ 
$X_{\nu ^{(n)}}^{R}=0$ \\ 
$X_{e^{(n)}}^{R}=-1$ \\ 
$X_{E^{(n)}}^{R}=-\frac{1}{2}-\frac{\sqrt{3}\beta }{2}$%
\end{tabular}
\\ \hline\hline
\begin{tabular}{c}
$\ell _{L}^{(n^{\ast })}=\left( 
\begin{array}{c}
e^{(n^{\ast })} \\ 
-\nu ^{(n^{\ast })} \\ 
E^{(n^{\ast })}%
\end{array}
\right) _{L}:\mathbf{3}^{\ast }$ \\ 
\\ 
$e_{R}^{(n^{\ast })}:\mathbf{1}$ \\ 
$\nu _{R}^{(n^{\ast })}:\mathbf{1}$ \\ 
$E_{R}^{(n^{\ast })}:\mathbf{1}$%
\end{tabular}
& 
\begin{tabular}{c}
$\left( 
\begin{array}{c}
-1 \\ 
0 \\ 
-\frac{1}{2}+\frac{\sqrt{3}\beta }{2}%
\end{array}
\right) $ \\ 
\\ 
$-1$ \\ 
$0$ \\ 
$-\frac{1}{2}+\frac{\sqrt{3}\beta }{2}$%
\end{tabular}
& 
\begin{tabular}{c}
\\ 
$X_{\ell ^{(n^{\ast })}}^{L}=\frac{1}{2}-\frac{\beta }{2\sqrt{3}}$ \\ 
\\ 
\\ 
$X_{e^{(n^{\ast })}}^{R}=-1$ \\ 
$X_{\nu ^{(n^{\ast })}}^{R}=0$ \\ 
$X_{E^{(n^{\ast })}}^{R}=-\frac{1}{2}+\frac{\sqrt{3}\beta }{2}$%
\end{tabular}
\\ \hline\hline
\end{tabular}%
\end{equation*}%
\end{center}
\caption{\textit{Fermionic content of }$SU\left( 3\right) _{L}\otimes
U\left( 1\right) _{X}$\textit{\ obtained by requiring only one lepton and
one quark }$SU\left( 3\right) _{L}$\textit{\ multiplet for each generation,
and no more than one right-handed singlet for each right-handed field. The
structure of left-handed multiplets is the one shown in Eq. (\protect\ref%
{notacion}). }$m$\textit{\ and }$n$\textit{\ label the quark and lepton
left-handed triplets respectively, while }$m^{\ast },n^{\ast }$\textit{\
label the antitriplets, see Eq. (\protect\ref{notacion}).}}
\label{tab:fercont}
\end{table}

It is possible to have in a single model any number of left-handed
multiplets in either the $\mathbf{3}$ or $\mathbf{3}^{\ast }$
representations. In the most general case, each multiplet can transform as 
\begin{equation}
\left\{ 
\begin{array}{l}
q_{L}^{\left( m\right) },q_{L}^{\left( m^{\ast }\right) }:\ m=\underset{3k\
triplets}{\underbrace{1,2,\ldots ,k}};\ m^{\ast }=\underset{3\left(
M-k\right) \ antitriplets}{\underbrace{k+1,k+2,\ldots ,M}} \\ 
\ell _{L}^{\left( n\right) },\ell _{L}^{\left( n^{\ast }\right) }\ \ \ \ :\
n=\underset{j\ triplets}{\underbrace{1,2,\ldots ,j}};\ n^{\ast }=\underset{%
N-j\ antitriplets}{\underbrace{j+1,j+2,\ldots ,N}}%
\end{array}%
\right.  \label{notacion}
\end{equation}%
where the first $3k$-th multiplets of quarks lie in the $\mathbf{3}$
representation while the latter $3\left( M-k\right) $ lie in the $\mathbf{3}%
^{\ast }$ representation for a total of $3M$ quark left-handed multiplets.
The factor $3$ in the number of quark left-handed multiplets owes to the
existence of three colors. Similarly the first $j$ left-handed multiplets of
leptons are taken in the representation $\mathbf{3}$ and the latter $\left(
N-j\right) $ are taken in the $\mathbf{3}^{\ast }$ representation, for a
total of $N$ leptonic left-handed multiplets. Table \ref{tab:jkrepres} shows
the possible multiplet structures in the fermionic sector compatible with
cancellation of anomalies with $\beta $ arbitrary \cite{331mirror}.
Representations with $N=1$ are forbidden.

Furthermore, the requirement for the model to be $SU\left( 3\right) _{c}$
vector-like demands the presence of right-handed quark singlets, while
right-handed neutral lepton singlets are optional. The possible charged
right-handed leptonic singlet structures compatible with cancellation of
anomalies are also studied in Ref. \cite{331mirror}. Tables \ref%
{tab:tetarest1}, \ref{tab:tetarest2} summarize the posssible leptonic
singlet structures and the solutions compatible with cancellation of
anomalies; in that tables we use $\Theta _{e^{\left( 1\right) }},\Theta
_{E^{\left( 1\right) }}$ to denote\ the number of singlets lying in the $%
\mathbf{3}$ representation associated with the ordinary and exotic leptons
repectively; while $\Theta _{e^{\left( j+1\right) }}\Theta _{E^{\left(
j+1\right) }}$stands for the corresponding singlets in the $\mathbf{3}^{\ast
}$ representation\footnote{%
It worths remarking that $\Theta _{l}=0,1$ because we have assumed that at
most one singlet is associated with each fermion multiplet.}.

\begin{table}[!tbp]
\begin{center}
\begin{tabular}{||c||c||}
\hline\hline
$N$ & Allowed representations \\ \hline\hline
2 & 
\begin{tabular}{|c|}
\hline
$\ell ^{(1)}:3$ \\ 
$\ell ^{(2)}:3^{\ast }$ \\ 
$q^{(1)}:3$ \\ 
$q^{(2)}:3^{\ast }$ \\ \hline
\end{tabular}
\\ \hline\hline
3 & 
\begin{tabular}{|l|}
\hline
$\ell ^{(1)},\ell ^{(2)},\ell ^{(3)}:3^{\ast }$ \\ 
$q^{(1)},q^{(2)}:3$ \\ 
$q^{(3)}:3^{\ast }$ \\ \hline
\end{tabular}
\begin{tabular}{|l|}
\hline
$\ell ^{(1)},\ell ^{(2)},\ell ^{(3)}:3$ \\ 
$q^{(3)}:3$ \\ 
$q^{(1)},q^{(2)}:3^{\ast }$ \\ \hline
\end{tabular}
\\ \hline\hline
4 & 
\begin{tabular}{|l|}
\hline
$\ell ^{(1)},\ell ^{(2)}:3$ \\ 
$\ell ^{(3)},\ell ^{(4)}:3^{\ast }$ \\ 
$q^{(1)},q^{(2)}:3$ \\ 
$q^{(3)},q^{(4)}:3^{\ast }$ \\ \hline
\end{tabular}
\\ \hline\hline
5 & 
\begin{tabular}{|l|}
\hline
$\ell ^{(5)}:3$ \\ 
$\ell ^{(1)},\ell ^{(2)},\ell ^{(3)},\ell ^{(4)}:3^{\ast }$ \\ 
$q^{(3)},q^{(4)},q^{(5)}:3$ \\ 
$q^{(1)},q^{(2)}:3^{\ast }$ \\ \hline
\end{tabular}
\begin{tabular}{|l|}
\hline
$\ell ^{(1)},\ell ^{(2)},\ell ^{(3)},\ell ^{(4)}:3$ \\ 
$\ell ^{(5)}:3^{\ast }$ \\ 
$q^{(1)},q^{(2)}:3$ \\ 
$q^{(3)},q^{(4)},q^{(5)}:3^{\ast }$ \\ \hline
\end{tabular}
\\ \hline\hline
6 & 
\begin{tabular}{c}
\begin{tabular}{|l|}
\hline
$%
\begin{tabular}{c}
$\ell ^{(1)},\ell ^{(2)},\ell ^{(3)},$ \\ 
$\ell ^{(4)},\ell ^{(5)},\ell ^{(6)}$%
\end{tabular}
\ :3^{\ast }$ \\ 
$q^{(1)},q^{(2)},q^{(5)},q^{(6)}:3$ \\ 
$q^{(3)},q^{(4)}:3^{\ast }$ \\ \hline
\end{tabular}
\begin{tabular}{|l|}
\hline
\begin{tabular}{c}
$\ell ^{(1)},\ell ^{(2)},\ell ^{(3)},$ \\ 
$\ell ^{(4)},\ell ^{(5)},\ell ^{(6)}$%
\end{tabular}
$:3$ \\ 
$q^{(3)},q^{(4)}:3$ \\ 
$q^{(1)},q^{(2)},q^{(5)},q^{(6)}:3^{\ast }$ \\ \hline
\end{tabular}
\\ 
\begin{tabular}{|l|}
\hline
$\ell ^{(1)},\ell ^{(2)},\ell ^{(3)}:3$ \\ 
$\ell ^{(4)},\ell ^{(5)},\ell ^{(6)}:3^{\ast }$ \\ 
$q^{(1)},q^{(2)},q^{(5)}:3$ \\ 
$q^{(3)},q^{(4)},q^{(6)}:3^{\ast }$ \\ \hline
\end{tabular}%
\end{tabular}
\\ \hline\hline
\end{tabular}%
\end{center}
\caption{\textit{Possible representations for the fermion left-handed
multiplets compatible with cancellation of anomalies. Each value of }$%
q^{\left( i\right) }$\textit{\ represents three left-handed quark multiplets
because of the color factor.}}
\label{tab:jkrepres}
\end{table}

\begin{table}[!tbp]
\begin{center}
\begin{tabular}{||c||c||c||c||c||}
\hline\hline
$\Theta _{e^{(1)}}$ & $\Theta _{E^{(1)}}$ & $\Theta _{e^{(j+1)}}$ & $\Theta
_{E^{(j+1)}}$ & Solution \\ \hline\hline
\ \ \thinspace 1\ \ \thinspace\  & \ \ \thinspace 0 \ \ \thinspace\  & \ \ \
\thinspace 0\ \ \thinspace\ \  & \ \ \ \thinspace \thinspace 1\ \ \
\thinspace \thinspace\  & $\quad \beta =-\sqrt{3}\quad $ \\ \hline\hline
0 & 1 & 1 & 0 & $\beta =\sqrt{3}$ \\ \hline\hline
\end{tabular}%
\end{center}
\caption{\textit{Solutions for} \textit{$N=2j=2k\geq 2$}}
\label{tab:tetarest1}
\end{table}
%-----------------------
%----------------------
\begin{table}[tbp]
\begin{center}
\begin{tabular}{||c||c||c||c||c||}
\hline\hline
$\Theta _{e^{(1)}}$ & $\Theta _{E^{(1)}}$ & $\Theta _{e^{(j+1)}}$ & $\Theta
_{E^{(j+1)}}$ & Solution \\ \hline\hline
0 & 0 & 0 & 0 & 
\begin{tabular}{c}
$\beta =\sqrt{3};$ $j=0$ \\ 
$\beta =-\sqrt{3};$ $j=N$%
\end{tabular}
\\ \hline\hline
0 & 0 & 1 & 1 & 
\begin{tabular}{c}
$\beta =-\sqrt{3};$ $\forall $ $j\neq 0$ \\ 
$\forall $ $\beta $; $j=0$%
\end{tabular}
\\ \hline\hline
1 & 1 & 0 & 0 & 
\begin{tabular}{c}
$\beta =\sqrt{3};$ $\forall $ $j\neq N$ \\ 
$\forall $ $\beta $; $j=N$%
\end{tabular}
\\ \hline\hline
0 & 1 & 1 & 1 & $j=0,\ \forall N,\ \beta =-1/\sqrt{3}$ \\ \hline\hline
1 & 1 & 0 & 1 & $j=N,\ \beta =1/\sqrt{3}$ \\ \hline\hline
1 & 1 & 1 & 1 & $%
\begin{array}{c}
\forall \beta ,\forall N,\ j\neq 0,N \\ 
\text{if\ }j=0\Rightarrow \beta =-\sqrt{3} \\ 
\text{if\ }j=N\Rightarrow \beta =\sqrt{3}%
\end{array}
$ \\ \hline\hline
\end{tabular}%
\end{center}
\caption{\textit{Solutions for} \textit{$N=\frac{j+3k}{2}\geq 2$, $0\leq
k\leq N$} }
\label{tab:tetarest2}
\end{table}

\vspace{-4mm}

\section{331 vector-like models with MF}

We can see from table \ref{tab:jkrepres} that models with $N=2,4$ are $%
SU\left( 3\right) _{L}\ $vector-like with respect to the quark and lepton
sectors separately. Besides, one of the three possible structures of
fermionic multiplets with $N=6$ is also vector-like in quark and leptons
sector. Taking the minimal fermionic content that could include the SM
fermions, we shall study the case of $N=4$. Finally, if we demand for the
model not to have exotic charges we are lead to only two values of the $%
\beta $ parameter i.e. $\beta =\pm 1/\sqrt{3}$. According to tables \ref%
{tab:tetarest1}, \ref{tab:tetarest2} we see that the only vector-like
structure with no exotic charges (i.e. $\beta =\pm 1/\sqrt{3}$) is the one
described by the first solution in the last row of table (\ref{tab:tetarest2}%
); in which exactly one right-handed singlet is associated with each
leptonic multiplet.

In particular, we shall examine the model with $N=4$ and $\beta =-1/\sqrt{3}$%
, where three families refer to the generations at low energies and the
other is a mirror family. This is a $SU\left( 3\right) _{L}\ $vector-like
model that has two multiplets in the $\mathbf{3}$ representation and two
multiplets in the $\mathbf{3}^{\ast }$ representation in both the quark and
lepton sectors. This extension of the 331 model is not reduced to the known
models with $\beta =-1/\sqrt{3}$ \cite{twelve}, because in such model the
leptons are in three 3-dimensional multiplets. From the phenomenological
point of view at low energies, the difference would be in generating ansatz
for the mass matrices in the lepton and quark sectors. As we mentioned
above, models with vector-like multiplets are necessary to explain the
family hierarchy, and mirror fermions are a possible source to solve the
deviation of the bottom quark asymmetries from the SM prediction.

\vspace{-4mm}

\section{Model with $N=4$ and $\protect\beta =-\frac{1}{\protect\sqrt{3}}$%
\label{modelN4}}

\begin{table}[!tbh]
\begin{center}
\begin{tabular}{||c||c||c||}
\hline\hline
$Quarks$ & $Q_{\psi }$ & $X_{\psi }$ \\ \hline\hline
$%
\begin{tabular}{c}
$q_{L}^{(m)}=\left( 
\begin{array}{c}
u^{(m)} \\ 
d^{(m)} \\ 
J^{(m)}%
\end{array}%
\right) _{L}:3$ \\ 
\\ 
$u_{R}^{(m)},$ $d_{R}^{(m)},$ $J_{R}^{(m)}:1$%
\end{tabular}%
\ \ $ & $%
\begin{tabular}{c}
$\left( 
\begin{array}{c}
\frac{2}{3} \\ 
-\frac{1}{3} \\ 
\frac{2}{3}%
\end{array}%
\right) $ \\ 
\\ 
$\frac{2}{3},-\frac{1}{3},\frac{2}{3}$%
\end{tabular}%
\ \ $ & $%
\begin{tabular}{c}
\\ 
$X_{q^{(m)}}^{L}=\frac{1}{3}$ \\ 
\\ 
\\ 
$X_{q^{(m)}}^{R}=Q_{q^{(m)}}$%
\end{tabular}%
\ \ $ \\ \hline\hline
$%
\begin{tabular}{c}
$q_{L}^{(3^{\ast })}=\left( 
\begin{array}{c}
d^{3^{\ast }} \\ 
-u^{3^{\ast }} \\ 
J^{3^{\ast }}%
\end{array}%
\right) _{L}:3^{\ast }$ \\ 
\\ 
$d_{R}^{3^{\ast }},$ $u_{R}^{3^{\ast }},$ $J_{R}^{3^{\ast }}:1$%
\end{tabular}%
\ \ $ & $%
\begin{tabular}{c}
$\left( 
\begin{array}{c}
-\frac{1}{3} \\ 
\frac{2}{3} \\ 
-\frac{1}{3}%
\end{array}%
\right) $ \\ 
\\ 
$-\frac{1}{3},\frac{2}{3},-\frac{1}{3}$%
\end{tabular}%
\ \ $ & $%
\begin{tabular}{c}
\\ 
$X_{q^{3^{\ast }}}^{L}=0$ \\ 
\\ 
\\ 
$X_{q^{3^{\ast }}}^{R}=Q_{q^{3^{\ast }}}$%
\end{tabular}%
\ \ $ \\ \hline\hline
$%
\begin{tabular}{c}
$q_{L}^{4^{\ast }}=\left( 
\begin{array}{c}
\widetilde{u}^{c} \\ 
\widetilde{d}^{c} \\ 
\widetilde{J}^{c}%
\end{array}%
\right) _{L}:3^{\ast }$ \\ 
\\ 
$\widetilde{u}_{R}^{c},$ $\widetilde{d}_{R}^{c},$ $\widetilde{J}_{R}^{c}:1$%
\end{tabular}%
\ \ $ & $%
\begin{tabular}{c}
$\left( 
\begin{array}{c}
-\frac{2}{3} \\ 
\frac{1}{3} \\ 
-\frac{2}{3}%
\end{array}%
\right) $ \\ 
\\ 
$-\frac{2}{3},\frac{1}{3},-\frac{2}{3}$%
\end{tabular}%
\ \ $ & $%
\begin{tabular}{c}
\\ 
$X_{q^{4^{\ast }}}^{L}=\frac{1}{3}$ \\ 
\\ 
\\ 
$X_{q^{4^{\ast }}}^{R}=Q_{q^{4^{\ast }}}$%
\end{tabular}%
\ \ $ \\ \hline\hline
$Leptons$ & $Q_{\psi }$ & $X_{\psi }$ \\ \hline\hline
\begin{tabular}{c}
$\ell _{L}^{(n)}=\left( 
\begin{array}{c}
\nu ^{(n)} \\ 
e^{(n)} \\ 
N^{0(n)}%
\end{array}%
\right) _{L}:3$ \\ 
\\ 
$\nu _{R}^{(n)},e_{R}^{(n)}:1$%
\end{tabular}
& 
\begin{tabular}{c}
$\left( 
\begin{array}{c}
0 \\ 
-1 \\ 
0%
\end{array}%
\right) $ \\ 
\\ 
$0,-1,0$%
\end{tabular}
& 
\begin{tabular}{c}
\\ 
$X_{\ell ^{(n)}}^{L}=-\frac{1}{3}$ \\ 
\\ 
\\ 
$X_{\ell ^{(n)}}^{R}=Q_{\ell ^{(n)}}$%
\end{tabular}
\\ \hline\hline
\begin{tabular}{c}
$\ell _{L}^{3^{\ast }}=\left( 
\begin{array}{c}
e^{3^{\ast }} \\ 
-\nu ^{3^{\ast }} \\ 
E^{3^{\ast }-}%
\end{array}%
\right) _{L}:3^{\ast }$ \\ 
\\ 
$e_{R}^{3^{\ast }},\nu _{R}^{3^{\ast }},E_{R}^{3^{\ast }-}:1$%
\end{tabular}
& 
\begin{tabular}{c}
$\left( 
\begin{array}{c}
-1 \\ 
0 \\ 
-1%
\end{array}%
\right) $ \\ 
\\ 
$-1,0,-1$%
\end{tabular}
& 
\begin{tabular}{c}
\\ 
$X_{\ell ^{3^{\ast }}}^{L}=\frac{2}{3}$ \\ 
\\ 
\\ 
$X_{\ell ^{3^{\ast }}}^{R}=Q_{\ell ^{3^{\ast }}}$%
\end{tabular}
\\ \hline\hline
\begin{tabular}{c}
$\ell _{L}^{4^{\ast }}=\left( 
\begin{array}{c}
\widetilde{\nu }^{c} \\ 
\widetilde{e}^{c} \\ 
\widetilde{N}^{0c}%
\end{array}%
\right) _{L}:3^{\ast }$ \\ 
\\ 
$\widetilde{\nu }_{R}^{c},\widetilde{e}_{R}^{c}:1$%
\end{tabular}
& 
\begin{tabular}{c}
$\left( 
\begin{array}{c}
0 \\ 
1 \\ 
0%
\end{array}%
\right) $ \\ 
\\ 
$0,1,0$%
\end{tabular}
& 
\begin{tabular}{c}
\\ 
$X_{\ell ^{4^{\ast }}}^{L}=-\frac{1}{3}$ \\ 
\\ 
\\ 
$X_{\ell ^{4^{\ast }}}^{R}=Q_{\ell ^{4^{\ast }}}$%
\end{tabular}
\\ \hline\hline
\end{tabular}%
\end{center}
\caption{\textit{Fermionic content of }$SU\left( 3\right) _{L}\otimes
U\left( 1\right) _{X}$\textit{\ , with }$\mathit{N}=4$, and $\mathit{m,n=1,2}
$\textit{. The 4th families which are in the }$\mathbf{3}^{\ast }\ $\textit{%
representation, are the mirror fermions of one of the families in the }$%
\mathbf{3}$ \textit{representation.}}
\label{fercont4}
\end{table}

We consider a model with $\beta =-1/\sqrt{3}$ which is similar to the model
described in Ref. \cite{twelve} at low energies due to the electromagnetic
charged assigned to different multiplets. However, this model is not the
same as the one in Ref. \cite{twelve} because the multiplets structure for
the quark sector is $SU(3)_{C}\otimes SU(3)_{L}$ vector-like, and the
leptonic part is not neccesary to cancel the quark anomalies. The leptonic
multiplets are also vector-like and anomaly free (see table \ref{fercont4}).
In the models described in the literature, the quarks anomalies are
cancelled out with the leptonic anomalies. In the model with $N=4$ and $%
\beta =-1/\sqrt{3}$ there are two $3$-multiplets for leptons and two $3$%
-multiplets for quarks and they generate the two heavy families of the SM.
Two $3^{\ast }$-multiplets for quarks and leptons correspond to the first SM
family; and the other two $3^{\ast }$, $q_{L}^{4\ast }$ and $l_{L}^{4\ast }$%
, correspond to a mirror fermion family of the third SM family. So with this
assignment, it is possible to get mixing between the bottom quark and its
mirror quark $d^{c}$ in order to modify the right-handed coupling of the
bottom quark with the $Z$ gauge boson which in turn might explain the
asymmetry deviations $A^{b}$ and $A_{FB}^{b}$ \cite{Chanowitz}. Such
discrepancy cannot be explained by a model with only left-handed multiplets
such as the SM \cite{anomalia} or the traditional 331 models \cite{Martinez}%
. The mixing in the mass matrix between the $b$ quark and its mirror fermion
permits a solution because the mirror couples with right-handed chirality to
the $Z_{\mu }$ gauge field of the SM. On the other hand, the mirror fermions
in the leptonic sector are useful to build up ansatz about mass matrices in
the neutrino and charged sectors. For the neutrinos corresponding to $%
SU\left( 2\right) _{L}$ doublets, right handed neutrino singlets are
introduced to generate masses of Dirac type.

As for the scalar spectrum, three types of representations are considered.
The three minimal triplets (whose VEV are shown in table \ref{tab:scalaralig}%
) that assure the spontaneous symmetry breaking (SSB)$\ 331\rightarrow
321\rightarrow 31$, and the masses for the gauge fields. Further, an
additional scalar in the adjoint representation is included. Such multiplet
permits a mixing of the mirror fermions with the ordinary fermions of the SM
in order to generate different ansatz for masses. The adjoint representation
acquires the VEV's displayed in table \ref{tab:scalaralig}. Finally, a
sextet representation can also be introduced as shown in table \ref%
{tab:scalaralig}, it acquires very small VEV's compared with the VEV's of
the 331 and electroweak scales $\nu _{\chi }$, $\nu _{\rho }$ and $\nu
_{\eta }$ since they belong to triplet components of $SU(2)_{L}$ and would
not break the relation for $\Delta \rho $. They also permit to generate
majorana masses for neutrinos.

\begin{table}[!tbh]
\begin{center}
\begin{tabular}{||l||l|l||}
\hline\hline
$\left\langle \chi \right\rangle _{0}$ & $\left( 
\begin{array}{ccc}
0 & 0 & \nu _{\chi }%
\end{array}%
\right) ^{T}$ & $X_{\chi }=-1/3$ \\ \hline
$\left\langle \rho \right\rangle _{0}$ & $\left( 
\begin{array}{ccc}
0 & \nu _{\rho } & 0%
\end{array}%
\right) ^{T}$ & $X_{\rho }=2/3$ \\ \hline
$\left\langle \eta \right\rangle _{0}$ & $\left( 
\begin{array}{ccc}
\nu _{\eta } & 0 & 0%
\end{array}%
\right) ^{T}$ & $X_{\eta }=2/3$ \\ \hline\hline
$\left\langle \phi \right\rangle _{0}$ & $\nu _{\chi }diag\left( 
\begin{array}{ccc}
1 & 1 & -2%
\end{array}%
\right) $ & $X_{\chi }=0$ \\ \hline\hline
$\left\langle S^{ij}\right\rangle _{0}$ & $V\left( 
\begin{array}{ccc}
1 & 0 & 0 \\ 
0 & 0 & 0 \\ 
0 & 0 & 1%
\end{array}%
\right) $ & $X_{S}=-1/3$ \\ \hline\hline
\end{tabular}%
\end{center}
\caption{\textit{Scalar sector with }$N=4\ $\textit{and its VEV's. $\protect%
\chi ,\protect\rho ,\protect\eta $ are triplets in the }$\mathit{\mathbf{3}}$
representation, $\protect\phi $ is a multiplet in the adjoint
representation, and $S$ lies in the sextet representation. \textit{$\protect%
\nu _{\protect\chi }$ is of the order of the first symmetry breaking. $%
\protect\nu _{\protect\rho }$, $\protect\nu _{\protect\eta }$ are of the
order of the electroweak scale. $V$ is much lower than the electroweak VEV.}}
\label{tab:scalaralig}
\end{table}

\vspace{-4mm}

\subsection{Mass matrix for quarks}

The Yukawa Lagrangian for quarks has the form

\begin{eqnarray}
\mathcal{L}_{Y}^{q} &=&\sum_{\Phi }\sum_{sing.}\sum_{m,m^{\prime
}=1}^{2}h_{q_{R}}^{m\varphi }\overline{q_{L}^{(m)}}q_{R}\Phi  \notag \\
&&+\frac{1}{2}\overline{q_{L}^{i(m)}}\left( q_{L}^{j(m^{\prime })}\right)
^{c}\left[ h_{\Phi }^{mm^{\prime }}\varepsilon ^{ijk}\Phi
_{k}+h_{S}^{mm^{\prime }}S^{ij}\right] +h_{q_{R}}^{3\Phi }\overline{%
q_{L}^{(3^{\ast })}}q_{R}\Phi ^{\ast }+h_{q_{R}}^{4\Phi }\overline{%
q_{L}^{(4^{\ast })}}q_{R}\Phi ^{\ast }  \notag \\
&&+\frac{1}{2}\overline{q_{iL}^{(3^{\ast })}}\left( q_{jL}^{(3^{\ast
})}\right) ^{c}\left[ Y_{\Phi }^{33}\varepsilon _{ijk}\Phi
^{k}+Y_{S}^{33}S_{ij}\right] +\frac{1}{2}\overline{q_{iL}^{(4^{\ast })}}%
\left( q_{jL}^{(4^{\ast })}\right) ^{c}\left[ Y_{\Phi }^{44}\varepsilon
_{ijk}\Phi ^{k}+h_{S}^{44}S_{ij}\right]  \notag \\
&&+\frac{1}{2}\overline{q_{iL}^{(3^{\ast })}}\left( q_{jL}^{(4^{\ast
})}\right) ^{c}\left[ Y_{\Phi }^{34}\varepsilon _{ijk}\Phi
^{k}+Y_{S}^{34}S_{ij}\right] +\frac{1}{2}\overline{q_{iL}^{(4^{\ast })}}%
\left( q_{jL}^{(3^{\ast })}\right) ^{c}\left[ Y_{\Phi }^{43}\varepsilon
_{ijk}\Phi ^{k}+Y_{S}^{43}S_{ij}\right]  \notag \\
&&+\frac{1}{2}h_{\phi }^{n3}\overline{q_{L}^{i(n)}}\left( q_{jL}^{(3^{\ast
})}\right) ^{c}\phi _{j}^{i}+\frac{1}{2}h_{\phi }^{3n}\overline{%
q_{iL}^{(3^{\ast })}}\left( q_{L}^{j(n)}\right) ^{c}\phi _{i}^{j}  \notag \\
&&+\frac{1}{2}h_{\phi }^{n4}\overline{q_{L}^{i(n)}}\left( q_{jL}^{(4^{\ast
})}\right) ^{c}\phi _{j}^{i}+\frac{1}{2}h_{\phi }^{4n}\overline{%
q_{iL}^{(4^{\ast })}}\left( q_{L}^{j(n)}\right) ^{c}\phi _{i}^{j}\text{ }%
+h.c,  \label{yukawa-q}
\end{eqnarray}%
with $\Phi $ being any of the $\eta $, $\rho ,$\ $\chi \ $multiplets, while $%
\phi ,\ $and $S$ correspond to the scalar adjoint and the sextet
representation of $SU(3)_{L}$ respectively. The third and fourth families
are written explicitly, since the fourth one correspond to a mirror fermion.
The constants $h_{\Phi }^{mm^{\prime }}$ and $Y_{\Phi }^{34}$ are
antisymmetric. It should be noted that all possible terms with scalar
triplets, adjoints, and sextets are involved. When we take the VEV's from
table \ref{tab:scalaralig}, the mass matrices are obtained.

For the mixing among up-type quarks in the basis $(u_{3^{\ast }},u_{1},u_{2},%
\widetilde{u},J_{1},J_{2},\widetilde{J})$ we get 
\begin{equation}
M^{up}=\left( 
\begin{array}{cc}
\mathcal{M}_{U} & \mathcal{M}_{JU} \\ 
\mathcal{M}_{UJ} & \mathcal{M}_{J}%
\end{array}%
\right) ,  \label{up-mass}
\end{equation}%
where 
\begin{equation*}
\mathcal{M}_{U}=\left( 
\begin{array}{cccc}
\nu _{\rho }h_{u_{3}}^{3\rho }\vspace*{0.2cm} & \nu _{\rho }h_{u_{1}}^{3\rho
} & \nu _{\rho }h_{u_{2}}^{3\rho } & h_{\chi }^{43}\nu _{\chi } \\ 
\nu _{\eta }h_{u_{3}}^{1\eta }\vspace*{0.2cm} & \nu _{\eta }h_{u_{1}}^{1\eta
} & \nu _{\eta }h_{u_{2}}^{1\eta } & h_{\phi }^{14}\nu _{\chi } \\ 
\nu _{\eta }h_{u_{3}}^{2\eta }\vspace*{0.2cm} & \nu _{\eta }h_{u_{1}}^{2\eta
} & \nu _{\eta }h_{u_{2}}^{2\eta } & h_{\phi }^{24}\nu _{\chi } \\ 
0 & 0 & 0 & \nu _{\eta }h_{\widetilde{u}}^{4\eta }%
\end{array}%
\right) \ ,\ \mathcal{M}_{J}=\left( 
\begin{array}{ccc}
\nu _{\chi }h_{J_{1}}^{1\varkappa }\vspace*{0.2cm} & \nu _{\chi
}h_{J_{2}}^{1\varkappa } & -2h_{\phi }^{14}\nu _{\chi } \\ 
\nu _{\chi }h_{J_{1}}^{2\varkappa }\vspace*{0.2cm} & \nu _{\chi
}h_{J_{2}}^{2\varkappa } & -2h_{\phi }^{24}\nu _{\chi } \\ 
0 & 0 & \nu _{\chi }h_{\widetilde{J}}^{4\varkappa }%
\end{array}%
\right) ,
\end{equation*}%
\begin{equation*}
\mathcal{M}_{UJ}=\left( 
\begin{array}{cccc}
\nu _{\chi }h_{u_{3}}^{1\varkappa }\vspace*{0.2cm} & \nu _{\chi
}h_{u_{1}}^{1\varkappa } & \nu _{\chi }h_{u_{2}}^{1\varkappa } & 0 \\ 
\nu _{\chi }h_{u_{3}}^{2\varkappa }\vspace*{0.2cm} & \nu _{\chi
}h_{u_{1}}^{2\varkappa } & \nu _{\chi }h_{u_{2}}^{2\varkappa } & 0 \\ 
0 & 0 & 0 & \nu _{\eta }h_{\widetilde{J}}^{4\eta }%
\end{array}%
\right) \ ,\ \mathcal{M}_{JU}=\left( 
\begin{array}{ccc}
\nu _{\rho }h_{J_{1}}^{3\rho }\vspace*{0.2cm} & \nu _{\rho }h_{J_{2}}^{3\rho
} & h_{\eta }^{34}\nu _{\eta _{1}} \\ 
\nu _{\eta _{1}}h_{J_{1}}^{1\eta }\vspace*{0.2cm} & \nu _{\eta
_{1}}h_{J_{2}}^{1\eta } & 0 \\ 
\nu _{\eta _{1}}h_{J_{1}}^{2\eta }\vspace*{0.2cm} & \nu _{\eta
_{1}}h_{J_{2}}^{2\eta } & 0 \\ 
0 & 0 & \nu _{\chi }h_{\widetilde{u}}^{4\chi }%
\end{array}%
\right)
\end{equation*}%
and $\left( u_{3^{\ast }},u_{1},u_{2}\right) $ correspond to the three
families of the SM, $\widetilde{u}$ refers to the mirror fermion of either $%
u_{1}$ or $u_{2}$, and $J_{1},J_{2},\widetilde{J}$ are the exotic quarks
with 2/3 electromagnetic charge.

For down-type quarks in the basis $(d_{3^{\ast }},d_{1},d_{2},\widetilde{d}%
,J_{3^{\ast }})$, the mass matrix yields 
\begin{equation}
M^{down}=\left( 
\begin{array}{ccccc}
\nu _{\eta }h_{d_{3}}^{3\eta }\vspace*{0.2cm} & \nu _{\eta }h_{d_{1}}^{3\eta
} & \nu _{\eta }h_{d_{2}}^{3\eta } & Y_{\chi }^{34}\nu _{\chi } & \nu _{\eta
}h_{J_{3}}^{3\eta } \\ 
\nu _{\rho }h_{d_{3}}^{1\rho }\vspace*{0.2cm} & \nu _{\rho }h_{d_{1}}^{1\rho
} & \nu _{\rho }h_{d_{2}}^{1\rho } & h_{\phi }^{14}\nu _{\chi } & \nu _{\rho
}h_{J_{3}}^{1\rho } \\ 
\nu _{\rho }h_{d_{3}}^{2\rho }\vspace*{0.2cm} & \nu _{\rho }h_{d_{1}}^{2\rho
} & \nu _{\rho }h_{d_{2}}^{2\rho } & h_{\phi }^{24}\nu _{\chi } & \nu _{\rho
}h_{J_{3}}^{2\rho } \\ 
0\vspace*{0.2cm} & 0 & 0 & \nu _{\rho }Y_{\widetilde{d}}^{4\rho } & 0 \\ 
\nu _{\chi }h_{d_{3}}^{3\varkappa } & \nu _{\chi }h_{d_{1}}^{3\varkappa } & 
\nu _{\chi }h_{d_{2}}^{3\varkappa } & Y_{\eta }^{43}\nu _{\eta } & \nu
_{\chi }h_{J_{3}}^{3\varkappa }%
\end{array}%
\right)  \label{down-mass}
\end{equation}%
$\left( d_{3^{\ast }},d_{1},d_{2}\right) $ are associated with the three SM
families, $\widetilde{d}$ is a down-type mirror quark of either $d_{1}$or $%
d_{2}$, and $J_{3^{\ast }}$ is an exotic down-type quark. When the adjoint
representation of the scalar fields is not taken into account, the mixing
between $q^{(m)}$ and the quark mirrors $q^{(4^{\ast })}$ does not appear.
Such mixing is important to change the right-handed coupling of the $b-$%
quark with the $Z_{\mu }$ gauge field, and look for a possible solution for
the deviation of the assymmetries $A_{b}$ and $A_{FB}^{b}$ of the SM with
respect to the experimental data. If the mixing with the mirror quarks were
withdrawn and the exotic particles were decoupled, the mirror quarks would
acquire masses of the order of the electroweak scale $\nu _{\rho }h_{%
\widetilde{d}}^{4\rho }$, $\nu _{\eta }h_{\widetilde{u}}^{4\rho }$ for the
up and down sectors, respectively.

\vspace{-4mm}

\subsection{Mass matrix for Leptons}

The Yukawa Lagrangian for leptons keeps the general form shown in Eq. (\ref%
{yukawa-q}) for the quarks. However, majorana terms could arise because of
the existence of neutral fields. By taking the whole spectrum including
right-handed neutrino singlets, Dirac terms are obtained for the charged
sector while Dirac and majorana terms appear in the neutral sector.

By including all the possible structures of VEV's, the charged sector in the
basis $(e_{3^{\ast }},e_{1},e_{2},\widetilde{e},E_{3^{\ast }}^{-})$ has the
following form

\begin{equation*}
M^{\ell \pm }=\left( 
\begin{array}{ccccc}
\nu _{\eta }h_{e_{3}}^{3\eta }\vspace*{0.2cm} & \nu _{\eta }h_{e_{1}}^{3\eta
} & \nu _{\eta }h_{e_{2}}^{3\eta } & h_{\chi }^{34}\nu _{\chi } & \nu _{\eta
}h_{J_{3}}^{3\eta } \\ 
\nu _{\rho }h_{e_{3}}^{1\rho }\vspace*{0.2cm} & \nu _{\rho }h_{e_{1}}^{1\rho
} & \nu _{\rho }h_{e_{2}}^{1\rho } & h_{\phi }^{14}\nu _{\chi } & \nu _{\rho
}h_{J_{3}}^{1\rho } \\ 
\nu _{\rho }h_{e_{3}}^{2\rho }\vspace*{0.2cm} & \nu _{\rho }h_{e_{1}}^{2\rho
} & \nu _{\rho }h_{e_{2}}^{2\rho } & h_{\phi }^{24}\nu _{\chi } & \nu _{\rho
}h_{J_{3}}^{2\rho } \\ 
0\vspace*{0.2cm} & 0 & 0 & \nu _{\rho }h_{\widetilde{e}}^{4\rho } & 0 \\ 
\nu _{\chi }h_{e_{3}}^{3\varkappa } & \nu _{\chi }h_{e_{1}}^{3\varkappa } & 
\nu _{\chi }h_{e_{2}}^{3\varkappa } & h_{\eta }^{43}\nu _{\eta _{1}} & \nu
_{\chi }h_{E_{3}}^{3\varkappa }%
\end{array}%
\right)
\end{equation*}%
the three first components $e_{i}\ $correspond to the ordinary leptons of
the SM, $\widetilde{e}$ is a mirror lepton of $e_{1}$ or $e_{2}$, and $%
E_{3^{\ast }}$ is an exotic lepton. Like in the case of the quark sector,
direct mixings are gotten between all the fields$\ \ell ^{(n)},\ell
^{3^{\ast }}$ and the mirrors $\ell ^{4^{\ast }}$ by means of the scalars $%
\chi ,\rho ,\eta $ and the adjoint $\phi $. The mass matrix of charged
leptons is similar to the mass matrix of the down-type quarks.

For the neutral lepton sector, we take the following basis of fields

\begin{eqnarray}
\psi _{L}^{0} &=&\left( \nu _{3L},\nu _{1L},\nu _{2L},\left( \widetilde{\nu }%
_{R}\right) ^{c},N_{1L}^{0},N_{2L}^{0},\left( \widetilde{N}_{R}^{0}\right)
^{c}\right) ^{T},  \notag \\
\psi _{R}^{0} &=&\left( \nu _{3R},\nu _{1R},\nu _{2R},\left( \widetilde{\nu }%
_{L}\right) ^{c}\right) ^{T},
\end{eqnarray}%
where $\nu _{iL}$ are the SM fields, $\nu _{iR}$ are sterile neutrinos and
the right-handed components of SM neutrinos. With these components the Dirac
mass matrix is constructed like the up quarks mass matrix; $\left( 
\widetilde{\nu }_{L,R}\right) ^{c}$ are mirror fermions, and $N_{iL}^{0}$
are exotic neutral fermions. The mass terms are written as%
\begin{equation}
\mathfrak{L}_{Y}^{0}=\left( \overline{\psi _{L}^{0}};\overline{\left( \psi
_{R}^{0}\right) ^{c}}\right) \left( 
\begin{tabular}{ll}
$M_{L}$ & $m_{D}$ \\ 
$m_{D}^{T}$ & $M_{R}$%
\end{tabular}%
\ \ \right) \left( 
\begin{array}{c}
\left( \psi _{L}^{0}\right) ^{c} \\ 
\psi _{R}^{0}%
\end{array}%
\right) +h.c,  \label{lepton-mass}
\end{equation}%
where very massive majorana terms $M_{R}$ have been introduced between the
singlets $\psi _{R}^{0}$, corresponding to sterile neutrinos with
right-handed chirality. We shall suppose that in this basis the mass matrix $%
M_{R}$ is diagonal. Such terms can be introduced without a SSB because they
are $SU(3)_{L}\otimes U(1)_{X}$ invariant. Besides, they correspond to heavy
majorana mass terms for the sterile heavy neutrinos. The majorana
contribution $M_{L}$ takes the form 
\begin{equation}
M_{L}=\frac{1}{2}\left( 
\begin{array}{cc}
\mathcal{M}_{\nu } & \mathcal{M}_{N\nu } \\ 
\mathcal{M}_{\nu N} & \mathcal{M}_{N}%
\end{array}%
\right) ,
\end{equation}%
where 
\begin{equation*}
\mathcal{M}_{\nu }=\left( 
\begin{array}{cccc}
0\vspace*{0.2cm} & 0 & 0 & -h_{\chi }^{34}\nu _{\chi } \\ 
0\vspace*{0.2cm} & Vh_{S}^{11} & Vh_{S}^{12} & h_{\phi }^{14}\nu _{\chi } \\ 
0\vspace*{0.2cm} & Vh_{S}^{21} & Vh_{S}^{22} & h_{\phi }^{24}\nu _{\chi } \\ 
h_{\chi }^{43}\nu _{\chi } & h_{\phi }^{14}\nu _{\chi } & h_{\phi }^{24}\nu
_{\chi } & Vh_{S}^{44}%
\end{array}%
\right) .
\end{equation*}%
The entries of the upper $3\times 3\ $submatrix correspond to majorana
masses for the ordinary neutrinos of the three SM families, which are
generated with the six dimensional representation of the scalar sector. If
such VEV were taken as null, or if we chose discrete symmetries to forbid
these terms, they can be generated through the see-saw mechanism of the form 
$m_{D}^{\dagger }M_{R}^{-1}m_{D}$. The other mass matrices are given by 
\begin{equation*}
\mathcal{M}_{N}=\left( 
\begin{array}{ccc}
Vh_{S}^{11}\vspace*{0.2cm} & Vh_{S}^{12} & -2h_{\phi }^{14}\nu _{\chi } \\ 
Vh_{S}^{21}\vspace*{0.2cm} & Vh_{S}^{22} & -2h_{\phi }^{24}\nu _{\chi } \\ 
-2h_{\phi }^{14}\nu _{\chi } & -2h_{\phi }^{24}\nu _{\chi } & Vh_{S}^{44}%
\end{array}%
\right) \ ,
\end{equation*}

\begin{equation*}
\mathcal{M}_{\nu N}=\left( 
\begin{array}{cccc}
0\vspace*{0.2cm} & h_{\rho }^{11}\nu _{\rho } & h_{\rho }^{12}\nu _{\rho } & 
0 \\ 
0\vspace*{0.2cm} & h_{\rho }^{21}\nu _{\rho } & h_{\rho }^{22}\nu _{\rho } & 
0 \\ 
-\nu _{\eta _{1}}h_{\eta }^{43} & 0 & 0 & h_{\rho }^{44}\nu _{\rho }%
\end{array}%
\right) \ ,\ \mathcal{M}_{N\nu }=\left( 
\begin{array}{ccc}
0\vspace*{0.2cm} & 0 & h_{\eta }^{34}\nu _{\eta _{1}} \\ 
-h_{\rho }^{11}\nu _{\rho }\vspace*{0.2cm} & -h_{\rho }^{12}\nu _{\rho } & 0
\\ 
-h_{\rho }^{21}\nu _{\rho }\vspace*{0.2cm} & -h_{\rho }^{22}\nu _{\rho } & 0
\\ 
0 & 0 & -h_{\rho }^{44}\nu _{\rho }%
\end{array}%
\right) .
\end{equation*}%
Where we have taken into account the VEV's of the scalar triplets $\chi
,\rho ,\eta $, the adjoint $\phi $ and the sextext $S$. The adjoint VEV's
ensure the direct mixings between $\ell ^{(n)}$ and the mirrors $\ell
^{(4^{\ast })}$. The Dirac terms of (\ref{lepton-mass}) are

\begin{equation}
m_{D}=\frac{1}{2}\left( 
\begin{array}{cccc}
\nu _{\rho }h_{\nu _{3}}^{3\rho }\vspace*{0.2cm} & \nu _{\rho }h_{\nu
_{1}}^{3\rho } & \nu _{\rho }h_{\nu _{2}}^{3\rho } & \nu _{\rho }h_{%
\widetilde{\nu }}^{3\rho } \\ 
\nu _{\eta }h_{\nu _{3}}^{1\eta }\vspace*{0.2cm} & \nu _{\eta }h_{\nu
_{1}}^{1\eta } & \nu _{\eta }h_{\nu _{2}}^{1\eta } & \nu _{\eta }h_{%
\widetilde{\nu }}^{1\eta } \\ 
\nu _{\eta }h_{\nu _{3}}^{2\eta }\vspace*{0.2cm} & \nu _{\eta }h_{\nu
_{1}}^{2\eta } & \nu _{\eta }h_{\nu _{2}}^{2\eta } & \nu _{\eta }h_{%
\widetilde{\nu }}^{2\eta } \\ 
\nu _{\eta }h_{\nu _{3}}^{4\eta } & \nu _{\eta }h_{\nu _{1}}^{4\eta } & \nu
_{\eta }h_{\nu _{2}}^{4\eta } & \nu _{\eta }h_{\widetilde{\nu }}^{4\eta } \\ 
\nu _{\chi }h_{\nu _{3}}^{1\chi }\vspace*{0.2cm} & \nu _{\chi }h_{\nu
_{1}}^{1\chi } & \nu _{\chi }h_{\nu _{2}}^{1\chi } & \nu _{\chi }h_{%
\widetilde{\nu }}^{1\chi } \\ 
\nu _{\chi }h_{\nu _{3}}^{2\chi }\vspace*{0.2cm} & \nu _{\chi }h_{\nu
_{1}}^{2\chi } & \nu _{\chi }h_{\nu _{2}}^{2\chi } & \nu _{\chi }h_{%
\widetilde{\nu }}^{2\chi } \\ 
\nu _{\chi }h_{\nu _{3}}^{4\chi } & \nu _{\chi }h_{\nu _{1}}^{4\chi } & \nu
_{\chi }h_{\nu _{2}}^{4\chi } & \nu _{\chi }h_{\widetilde{\nu }}^{4\chi }%
\end{array}%
\right) .  \label{mD}
\end{equation}

When the quarks and leptons spectra are compared (see table \ref{fercont4}),
it is observed that they are equivalent in the sense that both introduce the
same quantity of particles in the form of left-handed triplets and right
handed singlets (singlet components of neutrinos are taken). Nevertheless,
the Yukawa Lagrangian (and hence the mass matrices) of quarks and leptons
are not equivalent because the quarks have different values of the $X$
quantum number with respect to the leptons, this fact puts different
restrictions on the terms of both Yukawa Lagrangians.

In the limit $\nu _{\rho },\nu _{\eta }<<\nu _{\chi }$ and $V=0$, the
Physics beyond the SM could be decoupled at low energies leaving an
effective theory similar to a two Higgs doublet model (2HDM) with the
fermionic fields of the SM and the right-handed neutrinos that we introduced
in the particle content $\nu _{1R},\nu _{2R},\nu _{3R}$ to generate Dirac
type masses and be able to relate the neutrino sector with the up quark
sector. It allows to give a large mass to the up quark sector and the mass
pattern for the neutrinos. In this limit, the mass matrices that are
generated would be similar to the ansatz proposed in Ref. \cite{soniatwood}.
Considering the upper $3\times 3$ submatrix of $m_{D}$ in Eq. (\ref{mD}) and
imposing discrete symmetries, it can be written in the form 
\begin{equation}
m_{D}=\frac{1}{2}\left( 
\begin{array}{ccc}
\nu _{\rho }h_{\nu _{3}}^{3\rho }\vspace*{0.2cm} & \nu _{\rho }h_{\nu
_{1}}^{3\rho } & 0 \\ 
\nu _{\eta }h_{\nu _{3}}^{1\eta }\vspace*{0.2cm} & \nu _{\eta }h_{\nu
_{1}}^{1\eta } & \nu _{\eta }h_{\nu _{2}}^{1\eta } \\ 
0 & \nu _{\eta }h_{\nu _{1}}^{2\eta } & \nu _{\eta }h_{\nu _{2}}^{2\eta }%
\end{array}%
\right) .  \label{Dirac}
\end{equation}%
considering the same Yukawa couplings within each generation (i.e. the same $%
h_{\nu _{m}}^{n\Phi }$ for each pair $n\Phi $), we can write the matrix (\ref%
{Dirac}) as

\begin{equation}
m_{D}=\frac{\nu _{\eta }}{\sqrt{2}}\left( 
\begin{array}{ccc}
ct_{\beta }\vspace*{0.2cm} & ct_{\beta } & 0 \\ 
\delta b\vspace*{0.2cm} & b & b \\ 
0 & a & a%
\end{array}%
\right) ,
\end{equation}%
where $t_{\beta }\equiv \tan \beta =\frac{\nu _{\rho }}{\nu _{\eta }}$ is a
scalar mixing angle and $\delta $ is a real parameter that is fitted in
agreement with the neutrino oscillation data. If the third generation is $%
\nu _{3},$ the second is $\nu _{1}$ and the first is $\nu _{2},$ and taking $%
M_{R}=M_{diag}(\epsilon _{M3},\epsilon _{M2},\epsilon _{M1}),$ we obtain the
same mass ansatz and mixing as the Ref. \cite{soniatwood}. Thus, from the
see-saw mechanism we get

\begin{equation}
m_{\nu }=-m_{D}^{\dagger }M_{R}^{-1}m_{D}=m_{\nu }^{0}\left( 
\begin{array}{ccc}
\delta ^{2}\overline{\epsilon }+\omega \vspace*{0.2cm} & \delta \overline{%
\epsilon }+\omega & \delta \overline{\epsilon } \\ 
\delta \overline{\epsilon }+\omega \vspace*{0.2cm} & \epsilon +\omega & 
\epsilon \\ 
\delta \overline{\epsilon } & \epsilon & \epsilon%
\end{array}%
\right) ,
\end{equation}%
with $m_{\nu }^{0}=\frac{\nu _{\eta }^{2}}{2M},$ $\epsilon =\frac{a^{2}}{%
\epsilon _{M_{1}}}+\frac{b^{2}}{\epsilon _{M_{2}}},$ $\overline{\epsilon }=%
\frac{b^{2}}{\epsilon _{M_{2}}},$ $\omega =\frac{c^{2}t_{\beta }^{2}}{%
\epsilon _{M_{3}}},$ $\tan 2\theta _{23}\sim \frac{2r\omega }{\epsilon
\left( \delta ^{2}-r\right) },$ $\tan 2\theta _{12}\sim \frac{2g}{f},$ $%
\theta _{13}\sim \frac{\epsilon \left( \delta +r\right) }{2^{3/2}r\omega },$ 
$m_{1}\sim \epsilon m_{\nu }^{0}\left\{ 1-g\sin 2\theta _{12}+f\sin
^{2}\theta _{12}\right\} $ , $m_{2}\sim \epsilon m_{\nu }^{0}\left\{ 1+g\sin
2\theta _{12}+f\cos ^{2}\theta _{12}\right\} $, $m_{3}\sim 2\omega m_{\nu
}^{0},$ $r=\frac{\epsilon }{\overline{\epsilon }},$ $g=\frac{\left\vert
r-\delta \right\vert }{\sqrt{2}r},$ and $f=\frac{\delta ^{2}-2\delta -r}{2r}%
. $ As it is discussed in Ref. \cite{soniatwood}, if $m_{3}\sim \sqrt{\Delta
m_{atm}^{2}}$, $m_{2}\sim \sqrt{\Delta m_{sol}^{2}},$ and taking $t_{\beta }=%
\frac{\nu _{\rho }}{\nu _{\eta }}\gg O(1),$ it is possible to obtain a
natural fit for the observed neutrino hierarchical masses and mixing angles.
This result shows the good behavior of the model.

\vspace{-4mm}

\subsection{The mixing between the bottom quark and its mirror}

In order to look for a solution to the deviation from the $b$ asymmetries,
let us assume that the exotic quarks with charge $1/3$ acquire their mass in
the first SSB and that they are basically decoupled at electroweak energies.
On the other hand, let us suppose that the mass matrix of the three
generations of down quarks is approximately diagonal. In this way, the
mixing between the down quark of the third generation ($b$ quark) and its
corresponding mirror can be written as (see Eq. \ref{down-mass})%
\begin{eqnarray}
&&\left( 
\begin{array}{cc}
\bar{d}_{2} & \overline{\widetilde{d}}%
\end{array}%
\right) _{L}\ M\ \left( 
\begin{array}{c}
d_{2} \\ 
\widetilde{d}%
\end{array}%
\right) _{R}\ \ ,  \notag \\
M &\equiv &\left( 
\begin{array}{cc}
h_{d_{2}}^{2\rho }\nu _{\rho } & h_{\phi }^{24}\nu _{\chi } \\ 
0 & Y_{\widetilde{d}}^{4\rho }\nu _{\rho }%
\end{array}%
\right)  \label{defM}
\end{eqnarray}%
The eigenvalues of this mass matrix $M$, that correspond to the masses of
the $b$-quark and the mirror fermion are $h_{d_{2}}^{2\rho }\nu _{\rho }$
and $Y_{\widetilde{d}}^{4\rho }\nu _{\rho }$, respectively. To diagonalize
the mass matrix the following rotation is proposed%
\begin{equation}
\left( 
\begin{array}{c}
b \\ 
\widetilde{b}%
\end{array}%
\right) _{L(R)}=V_{L(R)}^{\dagger }\left( 
\begin{array}{c}
d_{2} \\ 
\widetilde{d}%
\end{array}%
\right) _{L(R)}
\end{equation}%
where $b\ $and $\widetilde{b}$ are the mass eigenstates for the bottom quark
and its mirror fermion respectively. $V_{L}$ and $V_{R}$ are $2\times 2$
matrices of rotation obtained from the matrices $MM^{\dagger }$ and $%
M^{\dagger }M$, respectively (see Eq. \ref{defM}). We shall assume that the
rotation angle of the left-handed quarks ($\theta _{L}$) is small enough,
since it would be tightly restricted by the electroweak processes. For the
right-handed angle we get%
\begin{equation}
\tan 2\theta _{R}=\frac{2h_{\phi }^{24}\nu _{\chi }Y_{\widetilde{d}}^{4\rho
}\nu _{\rho }}{(Y_{\widetilde{d}}^{4\rho }\nu _{\rho
})^{2}-(h_{d_{2}}^{2\rho }\nu _{\rho })^{2}-(h_{\phi }^{24}\nu _{\chi })^{2}}%
\approx \frac{2M_{Z^{\prime }}M_{F}}{M_{F}^{2}-M_{Z^{\prime }}^{2}}
\label{mezclar}
\end{equation}%
in the last line the $b$ quark mass was neglected and the VEV $\nu _{\chi }$
was approximated to$\ M_{Z^{\prime }}$.

When writing the neutral currents for $d_{2}$ and its mirror $\widetilde{d}$
we get 
\begin{eqnarray}
\mathcal{L}_{b}^{NC} &=&\frac{g}{2C_{W}}\overline{d_{2}}\gamma _{\mu }\left[
\left( 1-\frac{2}{3}S_{W}^{2}\right) P_{L}-\frac{2}{3}S_{W}^{2}P_{R}\right]
Z^{\mu }d_{2}  \notag \\
&+&\frac{g}{2C_{W}}\overline{\widetilde{d}}\gamma _{\mu }\left[ \left( 1-%
\frac{2}{3}S_{W}^{2}\right) P_{R}-\frac{2}{3}S_{W}^{2}P_{L}\right] Z^{\mu }%
\widetilde{d}
\end{eqnarray}%
After making the rotations for left and right-handed components of $d_{2}$,$%
\ \widetilde{d}$ quarks, and taking $\theta _{L}=0$, we can write the
right-handed current of the quark bottom mass eigenvalues as 
\begin{equation}
\frac{g}{2C_{W}}\overline{b}\gamma _{\mu }\left( \sin ^{2}\theta _{R}-\frac{2%
}{3}S_{W}^{2}\right) P_{R}Z^{\mu }b
\end{equation}%
and the electroweak right-handed coupling is modified by a factor 
\begin{equation}
\delta g_{R}=\sin ^{2}\theta _{R}
\end{equation}%
By making a combined fit for LEP and SLD measurements in terms of the left
and right currents of the $b$ quark, and substracting the central value of
the SM it is obtained that\ \cite{valencia2} 
\begin{equation}
\delta g_{R}=0.02
\end{equation}

It means that in order to solve the problem of the deviation of the anomaly $%
A_{b}$, it is necessary for the right-handed mixing angle to be of the order
of $\sin \theta _{R}\approx 0.1$. Replacing this value into Eq. (\ref%
{mezclar}) we find that $M_{Z^{\prime }}\approx 10M_{F}$. This is a
reasonable value if the mirror fermions lie at the electroweak scale and the
first breaking of the 331 model is of the order of the TeV scale.

\vspace{-4mm}

\section{Conclusions\label{conclusions}}

A general analysis of 331 models with $\beta $ arbitrary under the
assumption of minimal exotic spectrum, shows the possibility of finding
vector-like models with mirror fermions when the criterion of cancellation
of anomalies is taken. The existence of mirror fermions provides a possible
source to solve the problem of the deviation of the bottom quark
assymmetries from the SM predictions. On the other hand, the vector-like
nature of the model gives a possible solution to the fermion mass hierarchy
problem and in particular to the neutrino mass and mixing pattern. If we add
the assumption that no exotic charges are present in the model, the minimal
vector-like models that could contain the SM fermions are the ones with four
generations and $\beta =\pm 1/\sqrt{3}$.

In this manuscript we study the version with $N=4$ and $\beta =-1/\sqrt{3}$,
which is a vector-like model consisting of 3 triplets containing the SM
fermions plus one triplet containing mirror fermions of one of the SM
families. We choose the mirror fermions to be associated with the third
family of the SM. This $N=4\ $model is different from similar 331 versions
considered in the literature, and posseses strong phenomenological
motivations: the right-handed coupling of the $b-$quark with the $Z_{\mu }\ $%
gauge boson could be modified and may in turn explain the deviation of the $%
b $ asymmetries with respect to the SM prediction. In order to solve the $%
A_{b} $ puzzle, the right-handed mixing angle should be of the order of $%
\sin \theta _{R}\approx 0.1$, which in turn leads to $M_{Z^{\prime }}\approx
10M_{F}$ with $M_{Z^{\prime }}$ and $M_{F}$ denoting the masses for the
exotic neutral gauge boson and the mirror fermion respectively, this
relation is reasonable if $M_{F}$ lies at the electroweak scale and the
breaking of the 331 model lies at the TeV scale. On the other hand,
vector-like models are necessary to explain the family hierarchy. From the
phenomenological point of view, the model provides the possibility of
generating ansatz for masses at low energies in the quark and lepton sector.
It worths saying that the Physics beyond the SM could be decoupled at low
energies leaving an effective theory of two Higgs doublets with right-handed
neutrinos, and that the mass matrices generated are similar to the ansatz
proposed by Ref. \cite{soniatwood}. From such ansatz, a natural fit for the
neutrino hierarchical masses and mixing angles can be obtained.

Finally, we could find other possible 331 vector-like versions with mirror
fermions. For instance, we can analyze the model with $N=4$ but with the
mirror fermion associated with another SM family. Moreover, several
vector-like models with $N\geq 4$, with more mirror fermions could be
studied from phenomenological grounds (see table \ref{tab:jkrepres}). In
particular, we observe from table \ref{tab:jkrepres} that $N=6$ contains
models that are vector-like with respect to $SU\left( 3\right) _{L}$ in the
quark and lepton sectors.

\section*{Acknowledgments}

The authors acknowledge to Colciencias and Banco de la Rep\'{u}blica, for
the financial support. R. Martinez also acknowledge the kind hospitality of
Fermilab where part of this work was done.


\vspace{-4mm}

\begin{thebibliography}{99}
\bibitem{fourteen} Pisano, F. and V. Pleitez, Phys. Rev. \textbf{D46}, 410
(1992); P.H. Frampton, Phys. Rev. Lett. \textbf{69}, 2889 (1992); R. Foot,
O.F. Hernandez, F. Pisano and V. Pleitez, Phys. Rev. \textbf{D47}, 4158
(1993); P. H. Frampton, P. Krastev, J. T. Liu, Mod. Phys. Lett. \textbf{A9},
761 (1994); D. Ng, Phys. Rev. \textbf{D49}, 4805 (1994); L.A. S\'{a}nchez,
W.A. Ponce, and R. Mart\'{\i}nez, Phys. Rev. \textbf{D64}, 075013 (2001);
Nguyen Tuan Anh, Nguyen Anh Ky, Hoang Ngoc Long, Int. J. Mod. Phys. \textbf{%
A16}, 541 (2001); W.A. Ponce, J.B. Fl\'{o}rez and L.A. S\'{a}nchez, Int. J.
Mod. Phys. \textbf{A17}, 643 (2002); W.A. Ponce, Y. Giraldo, and L.A. S\'{a}%
nchez, Phys. Rev. \textbf{D67}, 075001 (2003).

\bibitem{Valle} E. D. Froggatt and H. B. Nielsen, Nucl. Phys. B147, 277
(1979); J. Schechter and J. W. F. Valle, Phys. Rev. \textbf{D22}, 2227
(1980); S. M. Barr, Phys. Rev. \textbf{D24}, 18951 (1981); S. M. Barr. Phys.
Rev. Lett. 92, 101601 (1994); Yosef Nir, Yael Shadmi, JHEP 0411 (2004) 055.

\bibitem{Martinez} G. Gonzalez-Sprinberg, R. Martinez, O. A. Sampayo, Phys.
Rev. \textbf{D71}, 115003 (2005).

\bibitem{Chanowitz} M. S. Chanowitz, Phys. Rev. Lett. 87, 231802 (2001); 
\emph{ibid. }Phys. Rev. \textbf{D60}, 073002 (2002).

\bibitem{331mirror} Rodolfo A. Diaz, R. Martinez, F. Ochoa [arXiv:
hep-ph/0411263]; to appear in Phys. Rev. D.

\bibitem{331us} Rodolfo A. Diaz, R. Martinez, F. Ochoa, Phys. Rev. \textbf{%
D69}, 095009 (2004); Rodolfo A. Diaz, R. Martinez, J. Mira, J.- Alexis
Rodriguez, Phys. Lett. \textbf{B552}, 287 (2003); R. Martinez, N. Poveda,
J.-Alexis Rodriguez [arXiv:hep-ph/0307147].

\bibitem{twelve} R. Foot, H.N. Long and T.A. Tran, Phys. Rev. \textbf{D50},
R34 (1994); H.N. Long, Phys. Rev. \textbf{D53}, 437 (1996); \textit{ibid}, 
\textbf{D54}, 4691 (1996); H.N. Long, Mod. Phys. Lett. \textbf{A13}, 1865
(1998).

\bibitem{anomalia} P. Bomert, C.P. Burgess, J.M. CLine, D. London and E.
Nardi, Phys.\ Rev.\ D\textbf{54}, 4275 (1996); G. Altarelli, F. Caravaglios,
G.F. Giudice, P. Gambino and G. Ridolfi, JHEP \textbf{06}, 018 (2001); X. He
and G. Valencia, Phys.\ Rev.\ D \textbf{66}, 013004 (2002) and Phys.\ Rev.\ D%
\textbf{68}, 033011 (2003).

\bibitem{soniatwood} David Atwood, Shaouly Bar-Shalom and Amarjit Soni,
[arXiv: hep-ph/0502234].

\bibitem{valencia2} J. Dress, [arXiv: hep-ex/0110077]; D. Abbaneo \textit{%
et. al.}, [arXiv: hep-ex/0112021]; Xiao-Gang He and G. Valencia, Phys.Rev. D%
\textbf{66} (2002) 013004; Erratum-ibid. D\textbf{66} (2002) 079901.
\end{thebibliography}
\end{document}